\documentclass[fleqn,12pt,twoside]{article}
\usepackage{espcrc1}

\usepackage{graphicx}

\newcommand{\AmS}{{\protect\the\textfont2
  A\kern-.1667em\lower.5ex\hbox{M}\kern-.125emS}}

\title{$\pi \eta $ pair hard  electroproduction and exotic hybrid mesons}

\author{I.V.~Anikin\address[LPT]{LPT, Universit{\'e} Paris-Sud,
                   91405-Orsay, France, UMR 8627 du CNRS}
         \address[BLTP]{BLTP, JINR,
           141980 Dubna, Russia }\,\address[CPHT]
          {CPHT, {\'E}cole
          Polytechnique, 91128 Palaiseau, France, UMR 7644 du CNRS }%
  ,
          B.~Pire\addressmark[CPHT],
          L.~Szymanowski\address
          {Soltan Institute for Nuclear Studies, Warsaw, Poland and \\
          Univ. de Li{\`e}ge, B-4000 Li{\`e}ge, Belgium },
          O.V.~Teryaev\addressmark[BLTP],
           S.~Wallon\addressmark[LPT]
          }

\begin{document}

\maketitle

\begin{abstract}
We show that hard electroproduction is
a promising way to study exotic hybrid mesons, in particular
through  the hybrid  decay channel $H \to \pi\eta $.
We discuss the $\pi\eta$ generalized distribution amplitude, calculate
the production amplitude and propose a forward-backward asymmetry
as a signal for the hybrid meson production.
\end{abstract}

\section{Introduction}
Present candidates for exotic  hybrid mesons with $J^{PC}=1^{-+}$ include
 $\pi_{1} (1400)$ which is
mostly seen through its $\pi\eta$ decay and $\pi_{1} (1600)$ which is seen through its
$\pi \eta' $ and $\pi \rho $ decays \cite{RPP}. The first experimental investigation of the hybrid with
$J^{PC}=1^{-+}$ as the resonance in
$\pi^-\eta$ mode was implemented by the Brookhaven collaboration E852 \cite{E852}.
Theoretically these states are objects of intense studies \cite{Close}, mostly through
lattice simulations \cite{Bernard}.
We recently studied exotic hybrid meson  electroproduction \cite{APSTWhyb} and showed that its cross section
is sizable in the kinematics of JLab or HERMES experiments.
We emphasize here that an angular asymmetry in the reaction
\begin{eqnarray}
\label{eN}
e(k_1)+N(p_1)\to e(k_2)+\pi^0(p_\pi)+\eta(p_\eta) + N(p_2)
\end{eqnarray}
will sign unambiguously the existence of the hybrid meson.

Exotic hybrid mesons are expected to be quite copiously electroproduced since
the normalization of its distribution amplitude has been shown to be quite similar to
 the one for the  $\rho$-meson. If an experiment is equipped with a
 recoil detector, the
hybrid production events may be identified through a missing mass reconstruction, and
all the decay channels may then be analyzed. If not, one will have to base
an identification process through the possible decay products of the hybrid meson $H$.
Since the hybrid candidate known as the $\pi_{1} (1400)$  has a dominant $\pi\eta $ decay
 mode, we  proceed to the description of the electroproduction  process (\ref{eN}) or
$ \gamma^*(q)+N(p_1)\to \pi^0(p_\pi)+\eta(p_\eta) + N(p_2)$.

\noindent
To perform a leading order computation of such process
we need to introduce  the concept of generalized distribution amplitude \cite{DGPT}.
Note that a very similar analysis may be carried for the $\pi \eta'$ decay mode of
the candidate $\pi_{1} (1600)$.

\section{$\pi\eta$ generalized distribution amplitude}

Let us briefly introduce and  discuss the generalized distribution amplitude
related to the $\pi\eta$--to--vacuum matrix element. On the basis of  Lorentz
invariance, the $\pi^0\eta$ GDA may be defined as :
\begin{eqnarray}
\label{hme2}
\langle \pi^0(p_\pi)\eta(p_\eta) |
\bar\psi_{f_2}(-\frac{z}{2})\gamma^{\mu}[-\frac{z}{2};\frac{z}{2}] \tau^3_{f_{2}f_{1}}
\psi_{f_1}(\frac{z}{2})|0\rangle=
p^{\mu}_{\pi\eta}\int\limits_{0}^{1}dy e^{i(\bar y-y)p_{\pi\eta}\cdot z/2}
\Phi^{(\pi\eta)}(y,\zeta, m_{\pi\eta}^2),
\end{eqnarray}
where  the total momentum of $\pi\eta$ pair is $p_{\pi\eta}=p_{\pi}+p_{\eta}$
and where $\tau^3$ is the usual Pauli matrix while $m^2_{\pi\eta}=p^2_{\pi\eta}$.
We omit here the $Q^2$ dependence of the $\pi^0\eta$ GDA's.
Note that the $\pi\eta$ distribution amplitude $\Phi^{(\pi\eta)}$ describes non resonant
as well as resonant contributions. It does not possess any symmetry properties
concerning the $\zeta$-parameter.

In the case of two different particles it is more convenient
to define the parameter $\zeta$ in the following way:
\begin{eqnarray}
\label{zeta}
&&\zeta=\frac{p_\pi^+}{(p_\pi+p_\eta)^+}-
\frac{m^2_\pi-m^2_\eta}{2m^2_{\pi\eta}},
\quad 1-\zeta=\frac{p_\eta^+}{(p_\pi+p_\eta)^+}+
\frac{m^2_\pi-m^2_\eta}{2m^2_{\pi\eta}}.
\end{eqnarray}
The relation between $\zeta$  and the angle $\theta_{cm}¥$ are
$2\zeta-1=\beta\cos\theta_{cm}\,,
\beta=2|{\bf p}|/m_{\pi\eta}$,
where ${\bf |p|}$ denote the modulus of three-dimension momentum of $\pi$ and $\eta$
mesons in the center--of--mass system.

In the reaction under  study, the $\pi\eta$ state may have
total momentum, parity and charge-conjugation in the following
sequence
$J^{PC}=0^{++},\, 1^{-+},\, 2^{++}, \, ...$,
that corresponds to the following values of the $\pi\eta$ orbital angular momentum $L$:
$L=0,\, 1, \, 2,\, ...,$
respectively. We can see that a resonance with a $\pi\eta$ decay mode for odd
orbital angular momentum $L$ should be considered as an exotic meson.

The mass region around $1400$ ${\rm MeV}$ is dominated by the strong $a_2(1329)\,(2^{++})$
resonance \cite{Adams}. It is therefore natural to look for the
interference of the amplitudes of hybrid and $a_{2}$ production, which is linear, rather
than quadratic in the hybrid electroproduction amplitude.
Such interference arises from the usual representation of the $\pi\eta$ generalized
distribution amplitude
in the form suggested by its asymptotic expression :
\begin{eqnarray}
\label{asPhi}
\Phi^{(\pi\eta),\,a}(y,\zeta, m_{\pi\eta}^2)=10 y (1-y)
C^{(3/2)}_1(2y-1)
\sum_{l=0}^{2} B_{1l}(m_{\pi\eta}^2) P_l(\cos\theta).
\end{eqnarray}
Keeping only $L=1$ and $L=2$ terms, and using the description of tensor
meson distribution amplitudes suggested by Ref \cite{BraunKivel},
we model the $\pi\eta$ distribution amplitude in the following form:
\begin{eqnarray}
\label{approx}
\Phi^{(\pi\eta)}(y,\zeta, m_{\pi\eta}^2)=30 y (1-y)(2y-1)
\biggl[
\,B_{11}(m_{\pi\eta}^2) P_1(\cos\theta) +
B_{12}(m_{\pi\eta}^2)
P_2(\cos\theta)
\biggr],
\end{eqnarray}
with the coefficient functions $B_{11}(m_{\pi\eta}^2)$ and
$B_{12}(m_{\pi\eta}^2)$   related to corresponding
Breit-Wigner amplitudes when
$m^2_{\pi\eta}$ is in the vicinity of $ M^2_{a_2},\,M^2_H$. We have
(see the technical details in Ref \cite{APSTWhyb}):
\begin{eqnarray}
\label{B112}
 B_{11}(m^2_{\pi\eta})=
\frac{5}{3}\,
\frac{g_{H\pi\eta}f_H M_H \beta}{M^2_H-m^2_{\pi\eta}-i\Gamma_H M_H},
\quad
B_{12}(m^2_{\pi\eta})=
\frac{10}{9} \frac{i g_{a_2\pi\eta} f_{a_2} M^2_{a_2} \beta^2}
{M^2_{a_2}-m^2_{\pi\eta}-i\Gamma_{a_2} M_{a_2}},
\end{eqnarray}
where $f_H$, $f_{a_2}$, $g_{H\pi\eta}$ and $g_{a_2\pi\eta}$ are the corresponding coupling constants.

\section{Differential cross section for $\pi\eta$ electroproduction}

The amplitude of reaction (\ref{eN}):
\begin{eqnarray}
\label{genam}
T^{\pi^0\eta}=\bar u(k_2,s_2)\gamma\cdot\varepsilon_L u(k_1,s_1)
\frac{1}{q^2}{\cal A}_{(q)}^{\pi^0\eta },
\quad
|T^{\pi^0\eta}|^2=
\frac{4e^2(1-y_l)}{Q^2 y_l^2}
|{\cal A}_{(q)}^{\pi^0\eta }|^2\,,
\end{eqnarray}
where
\begin{eqnarray}
\label{qdsim3}
&&{\cal A}_{(q)}^{\pi^0\eta }=
\frac{e\pi\alpha_s C_F}{N_c\,Q}
\biggl[ e_u {\cal H}_{uu} -e_d {\cal H}_{dd}\biggr]
\biggl[ B_{11}(m_{\pi\eta}^2) P_1(\cos\theta_{cm}) +
B_{12}(m_{\pi\eta}^2)P_2(\cos\theta_{cm})
\biggr].
\end{eqnarray}
Finally, the differential cross section of process (\ref{eN})  takes
the form
\begin{eqnarray}
\label{xsec}
\frac{d\sigma^{\pi^0\eta}}{dQ^2\, dy_l\,d\hat t\,
dm_{\pi\eta}\, d(\cos\theta_{cm})}
=\frac{1}{4(4\pi)^5}\,
\frac{m_{\pi\eta}\beta}{y_l\lambda^2(\hat s,-Q^2,m^2_N)}
\,|T^{\pi^0\eta}|^2.
\end{eqnarray}

\section{Angular asymmetry}

\begin{figure}
$$\rotatebox{270}{\includegraphics[width=6cm]{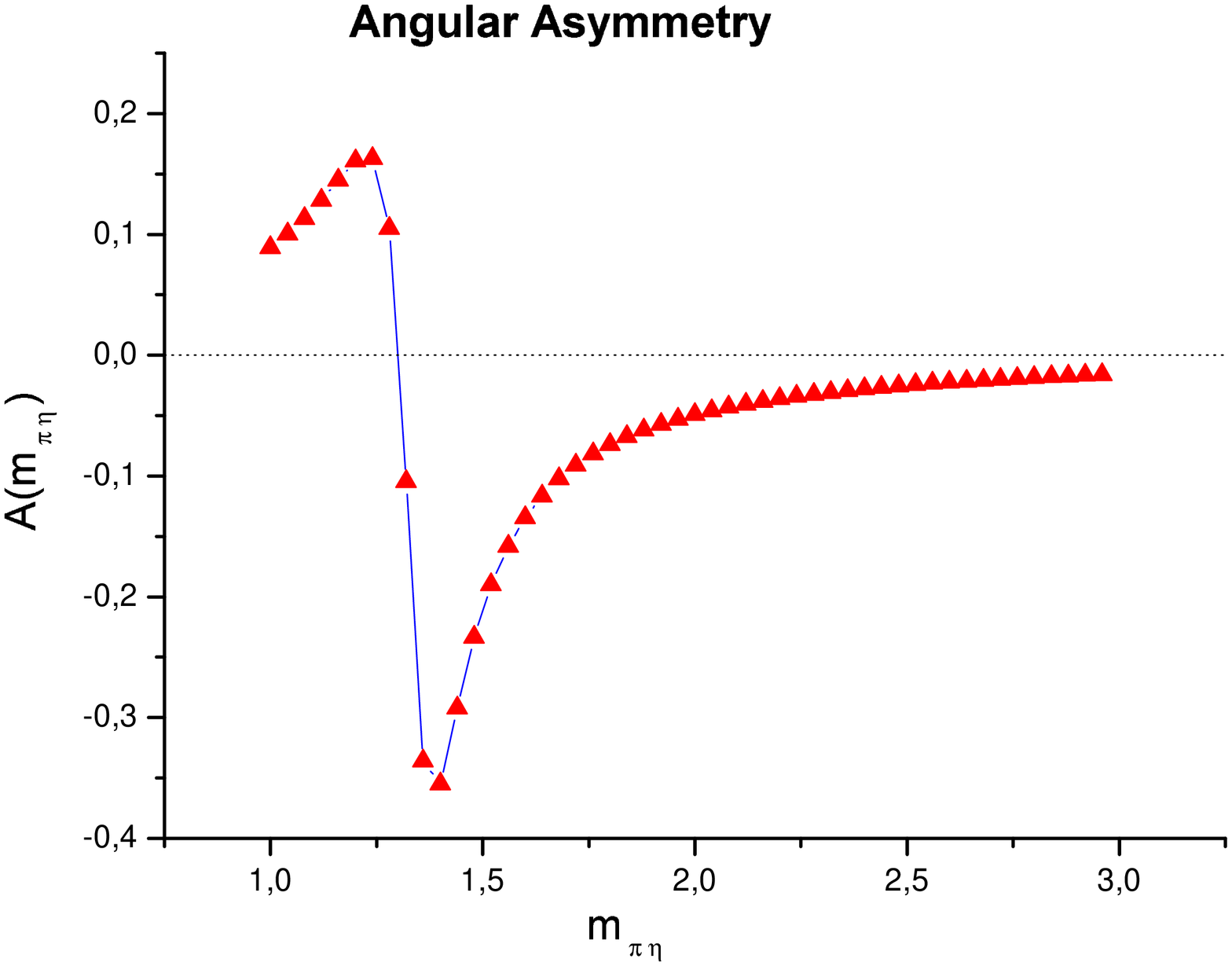}}$$
\caption{The angular asymmetry as a function of $m_{\pi\eta}$.}
\label{angular}
\end{figure}

Asymmetries are often a good way to get a measurable signal for a small
amplitude, by taking profit of its interference with a larger one. In our case,
since the hybrid production amplitude may be rather small with respect to
a continuous background, we propose to use the supposedly
large amplitude for $a_{2}$ electroproduction as a magnifying lens to unravel
the presence of the exotic hybrid meson. Since these two amplitudes describe
different orbital angular momentum of the $\pi$ and $\eta$ mesons, the asymmetry
which is sensitive to their interference is an angular asymmetry  defined by
\begin{eqnarray}
\label{anas}
 A(Q^2, y_l,\hat t, m_{\pi\eta})=
 \frac{\int \cos\theta_{cm} \,
d\sigma^{\pi^0\eta}(Q^2, y_l,\hat t, m_{\pi\eta}, \cos\theta_{cm} )}{
\int d\sigma^{\pi^0\eta}
(Q^2, y_l,\hat t, m_{\pi\eta}, \cos\theta_{cm} )}
\end{eqnarray}
as a weighted integral over polar angle $\theta_{cm}$ of the relative momentum of
$\pi$ and $\eta$ mesons.

Note that this angular asymmetry  is completely similar to the
charge asymmetry which was studied in $\pi^+\pi^-$ electroproduction
at HERMES \cite{hermes} and discussed in Ref \cite{HPST}.

Due to the fact that the $\cos\theta_{cm}$-independent factors in both
the numerator and denominator of (\ref{anas}) are completely factorized and,
on the other hand, these factors are the same, we are able to rewrite the asymmetry (\ref{anas})
as
\begin{eqnarray}
\label{anas2}
 A(m_{\pi\eta})=
\frac{\int d(\cos\theta_{cm}) \, \cos\theta_{cm}
\biggl| B_{11}(m_{\pi\eta}^2) P_1(\cos\theta_{cm}) +
B_{12}(m_{\pi\eta}^2)P_2(\cos\theta_{cm})
\biggr|^2}{
\int d(\cos\theta_{cm}) \,
\biggl| B_{11}(m_{\pi\eta}^2) P_1(\cos\theta_{cm}) +
B_{12}(m_{\pi\eta}^2)P_2(\cos\theta_{cm})
\biggr|^2}.
\end{eqnarray}
Our estimation of the asymmetry (\ref{anas2}) is shown on Fig.\ref{angular}.
Since the numerator of (\ref{anas2}),
{\it i.e.} the real part of the product of $B_{11}(m_{\pi\eta}^2)$ and
$B_{12}^*(m_{\pi\eta}^2)$, is proportional to the cosine of the phase difference
$\Delta\delta_{1,2}=\delta_{l=1}-\delta_{l=2}$, the zeroth
value of (\ref{anas2})  takes place at $\Delta\delta_{1,2}=\pi/2$.
This is achieved for $m_{\pi\eta}\approx 1.3 \, {\rm GeV}$.
Besides, one can see from Fig. \ref{angular} that the first positive extremum
is located at $m_{\pi\eta}$ around the mass of $a_2$ meson while the second negative extremum
corresponds to the hybrid meson mass.

\vspace*{.1cm} \noindent
{\bf Acknowledgments}

\vspace*{.1cm} \noindent
This work is supported in part by
and RFBR (Grant 03-02-16816) and NATO Grant, by
the Joint Research Activity "Generalised
Parton Distributions" in I3HP Project
of the European Union, contract No.~RII3-CT-2004-506078 and by the
Polish Grant 1 P03B 028 28.
L.Sz.\ is a Visiting Fellow of the Fonds National pour la Recherche
Scientifique (Belgium).  The work of B.P. and L. Sz.\ is partially
supported by the French-Polish scientific agreement Polonium.


\begin{thebibliography}{9}
\bibitem{RPP}
S. Eidelman et al, Phys.\ Lett.\ {\bf B592} (2004) 1;
C.~Amsler and N.~A.~Tornqvist,
Phys.\ Rept.\  {\bf 389} (2004) 61.

\bibitem{E852}
D.~R.~Thompson {\it et al.}  [E852 Collaboration],
Phys.\ Rev.\ Lett.\  {\bf 79} (1997) 1630;


\bibitem{Close}
F.~E.~Close and P.~R.~Page,
Phys.\ Rev.\ D {\bf 52} (1995) 1706.

\bibitem{Bernard}
C.~Bernard {\it et al.},
Phys.\ Rev.\ D {\bf 68} (2003) 074505.

\bibitem{APSTWhyb}
I.~V.~Anikin {\it et al},
Phys.\ Rev.\ D {\bf 70}, 011501 (2004) and arXiv:hep-ph/0411407.

\bibitem{DGPT} M.~Diehl {\it et al.},
Phys.\ Rev.\ Lett.\  {\bf 81} (1998) 1782
and arXiv:hep-ph/9901233;
M.~V.~Polyakov, Nucl.\ Phys.\   {\bf B 555} (1999) 231;
B.~Lehmann-Dronke {\it et al},
Phys.\ Lett.\  {\bf B 475} (2000) 147;
B.~Pire and L.~Szymanowski,
Phys.\ Lett.\ B {\bf 556} (2003) 129.

\bibitem{Adams}
G.~S.~Adams {\it et al.}  [E852 Collaboration],
Phys.\ Rev.\ Lett.\  {\bf 81} (1998) 5760 .

\bibitem{BraunKivel}
V.~M.~Braun and N.~Kivel,
Phys.\ Lett.\ B {\bf 501} (2001) 48.


\bibitem{hermes}
A.~Airapetian {\it et al.}  [HERMES Collaboration],
Phys.\ Lett.\ B {\bf 599} (2004) 212
%


\bibitem{HPST}
B.~Lehmann-Dronke {\it et al.}
Phys.\ Rev.\ D {\bf 63} (2001) 114001;
 P.~H{\"a}gler {\it et al.}
Phys.\ Lett.\ B {\bf 535} (2002) 117 and
Eur.\ Phys.\ J.\ C {\bf 26} (2002) 261.



\end{thebibliography}
\end{document}